\documentclass[prd,showpacs,nofootinbib,preprintnumbers]{revtex4}
\usepackage{amsmath}
\usepackage{amsfonts}
\usepackage{graphicx}
\usepackage{dcolumn}
\usepackage{hyperref}

\def\be{\begin{equation}}
\def\ee{\end{equation}}
\def\bea{\begin{eqnarray}}
\def\eea{\end{eqnarray}}

\def\5{\overline 5}

\begin{document}

 \title{Catalyst type of interactions between dark energy and dark matter}
\author{Yan-Hong Guo}
\email{guoyh2015@lzu.edu.cn}
\affiliation{Institute of Theoretical Physics, Lanzhou University, Lanzhou 730000, China}

\author{Zhong-Xi Yu}
\email{yuzhx14@lzu.edu.cn}
\affiliation{Institute of Theoretical Physics, Lanzhou University, Lanzhou 730000, China}

\author{Ji-Rong Ren\footnote{Corresponding author}}
\email{renjr@lzu.edu.cn}
\affiliation{Institute of Theoretical Physics, Lanzhou University, Lanzhou 730000, China}

\begin{abstract}
In this paper, we focus on three specific interactions of dark sector in the existence of baryonic matter and radiation. First, we  attempt to assume baryonic matter and radiation can affect the conversion between dark energy and dark matter like the way catalyst influences the conversion rate of two materials in some reversible chemical reactions. Then we present phase space analysis for every special interaction model. Finally, for every case, we obtain a stable attractor solution that can alleviate the coincidence problem.

\end{abstract}
\maketitle
\section{Introduction}
A lot of observational data such as Supernova Ia\cite{price2,price1}, large scale structure\cite{price4,price3}, the baryon acoustic oscillations\cite{price5}and cosmic microwave background radiation\cite{price8,price9,price6,price7}, suggests that our Universe is experiencing accelerating process. Within the framework of General Relativity, the accelerated expansion is caused by a exotic energy
 density component which has to possess a huge negative pressure, called dark energy. However, it is still not clear what dark energy actually is.

The cosmological constant $\Lambda$ which has a constant state parameter $\omega =-1$ is the simplest dark energy candidate. In the standard $\Lambda$-cold dark matter ($\Lambda$ CDM) model of the Universe, the dark energy is regarded as the constant vacuum energy and the cold dark matter only interacts with other components gravitationally. ``Quintessence'' which is a canonical scalar field is also a viable candidate \cite{price12,price13,price10,price11}, Besides,there are other scalar-field dark energy models, such as tachyon\cite{paper7}, phantom\cite{paper6}, quintom\cite{paper8}, ghost condensates\cite{paper9}, etc. Though there are so many various models to describe dark energy, until now, $\Lambda$-cosmology is still the best cosmological model. However, this model is troubled with the so-called fine-tuning problem and coincidence problem\cite{price15,price16}.

 A new method that might handle the mentioned problem is considering the interaction of dark sector. In recent years, The interacting dark energy models have been studied by several papers\cite{interaction8,interaction9,price22,interaction3,interaction4,interaction1,interaction5,interaction2,interaction6,interaction7}. The research of phase space analysis is the one compellent test for dark energy models. Particularly, the attractor solutions are independent of a wide range of initial conditions. If the dark energy models have $ \frac{\Omega _{\text{DE}}}{\Omega _{\text{DM}}}$ of the order one and an accelerated stable attractor solution, then the coincidence problem can be alleviated.

 In our work, we also focus on the interaction between dark energy and dark matter in the existence of baryonic matter and radiation. Based on the fact that we are not exactly sure what connection they have among the four components, so we attempt to give a new interesting relationship between normal matter (baryonic and radiation) and dark sector. Just like the action of catalyst in some reversible chemical reactions, although they could change the reaction process, they can not change the balance point of the reaction and themselves are always conserved. We tentatively consider baryonic or radiation or both of them as the ``catalyst'' of dark sector, in other words, we assume the baryonic matter and radiation are always conserved, however, they could affect the interaction process of dark sector and their influences are embodied in the interaction form Q. Specially, we mainly discuss three relatively simple and natural interaction forms $Q=\alpha H(\rho_{d}+\rho_{m}+\rho_{b})$, $Q=\beta H(\rho_{d}+\rho_{m}+\rho_{r})$ and $Q=\gamma H(\rho_{d}+\rho_{m}+\rho_{b}+\rho_{r})$. Fortunately, for every interaction, we could obtain a stable attractor solution which can describe the current accelerated expansion of the Universe and alleviate the coincidence problem.

  This paper is organized as follows: in section II we establish the interacting cosmological
 framework and construct an autonomous dynamical system. In section III-V we discuss phase space analysis and analyze the stability of critical points for three interacting cosmological models. Finally, we summarize our
 results and conclusions in section VI.

\section{ dynamical system of interacting  Dark Energy Model  }
In this work, we consider a flat FRW Universe. The Friedmann and Raychaudhuri equations are given by
\begin{equation}
H^{2}=\frac{\kappa^{2}}{3}\rho=\frac{\kappa^{2}}{3}(\rho_{d}+\rho_{m}+\rho_{b}+\rho_{r}),
\end{equation}
\begin{eqnarray}
\dot{H}&=&-\frac{\kappa^{2}}{2}(\rho+p)\nonumber\\
&=&-\frac{\kappa^{2}}{2}((1+w_{d})\rho_{d}+\rho_{m}+\rho_{b}+\frac{4}{3}\rho_{r}).
\end{eqnarray}
where $\kappa^{2}\equiv8\pi G$.
In our scenario, the Universe contains dark energy, cold dark matter, baryonic matter and radiation.
The total energy conservation equation is
\begin{equation}
\dot{\rho}+3H(\rho+p)=0,
\end{equation}
where $H$ is the Hubble parameter, $\rho$ is the total energy
density and $p$ is the total pressure of the four components. Furthermore,we consider dark energy can interact with dark matter through Q. baryonic matter and radiation are always conserved respectively. Then the conservation
equation is written as
\begin{equation}
\dot{\rho}_{b}+3H\rho_{b}=0,
\end{equation}
\begin{equation}
\dot{\rho}_{r}+4H\rho_{r}=0,
\end{equation}
\begin{equation}
\dot{\rho}_{d}+3H(1+w_{d})\rho_{d}=-Q,
\end{equation}
\begin{equation}
\dot{\rho}_{m}+3H\rho_{m}=Q,
\end{equation}
where the subscripts $b$, $r$, $d$ and $m$ respectively denote baryonic matter, radiation, dark energy and cold dark matter. Here we consider the simplest case of dark energy which has a constant state parameter $w_{d}=p_{d}/\rho_{d}$~\cite{constant} even though the state parameter for dark energy could also be dynamic. There is as yet no basis fundamental theory for a special coupling of dark sector because of the unknown nature of dark sector. So the interaction term $Q$ have to be chosen in a phenomenological way~\cite{phys40}. Moreover, new observations support that we should increase the number of different possible forms of Q. In this paper, we focus on three relatively simple and natural interaction forms, namely $Q=\alpha H(\rho_{d}+\rho_{m}+\rho_{b})$, $Q=\beta H(\rho_{d}+\rho_{m}+\rho_{r})$ and $Q=\gamma H(\rho_{d}+\rho_{m}+\rho_{b}+\rho_{r})$. $Q$ represents the energy density exchange between dark energy and dark matter and the sign of $Q$ reflect the direction of energy transfer. A positive $Q$ corresponds transfer of energy occurs from dark energy to dark matter, while a negative $Q$ represents the opposite direction.

To analyze the evolution of the dynamical system, we introduce the following dimensionless variables:
\begin{equation}
x\equiv\frac{\kappa^{2}\rho_{d}}{3H^{2}},~~~~~~~~y\equiv\frac{\kappa^{2}\rho_{m}}{3H^{2}},~~~~~~~~z\equiv\frac{\kappa^{2}\rho_{b}}{3H^{2}},~~~~~~~~v\equiv\frac{\kappa^{2}\rho_{r}}{3H^{2}},~~~~~~~~u\equiv\frac{\kappa^{2}Q}{3H^{3}}.
\end{equation}
Then, from Eq.(1) and eq.(2), we have
\begin{equation}
-\frac{\dot{H}}{H^{2}}=\frac{3}{2}(1+w_{d}x+\frac{1}{3}v).
\end{equation}
and the Friedmann constraint is
\begin{equation}
x+y+z+v=1,
\end{equation}
Furthermore, using these dimensionless variables, the total state parameter of the four components is given by
\begin{equation}
w=\frac{p}{\rho}=\frac{w_{d}x+\frac{1}{3}v}{x+y+z+v}=w_{d}x+\frac{1}{3}v.
\end{equation}
inserting the expression (8) into Eqs.(4)-(7), we can obtain the following autonomous system:
\begin{eqnarray}
x^{'}&=&-3w_{d}x(1-x)-u+x v,\\
y^{'}&=&3w_{d}x y+u+y v,\\
z^{'}&=&3w_{d}x z+z v,\\
v^{'}&=&3w_{d}x v+v^{2}-v,
\end{eqnarray}
where the prime denotes a derivative with respect to $N\equiv \ln a$. By setting $x^{'}=y^{'}=z^{'}=v^{'}=0$ and $(x+y+z+v=1)$, we can
find the general solution of the critical points $(x_{c},y_{c},z_{c},v_{c})$ of the autonomous system (12)-(15). In order to determine the linear stability of these points, one must analyse the following Jacobian matrix of partial derivatives
\\
\begin{center}
M=$\left(
\begin{array}{cccc}

$$\frac{\partial x'}{\partial x}$$ & $$\frac{\partial x'}{\partial y}$$ & $$\frac{\partial x'}{\partial z}$$ & $$\frac{\partial x'}{\partial v}$$ \\
 $$\frac{\partial y'}{\partial x}$$ & $$\frac{\partial y'}{\partial y}$$ & $$\frac{\partial y'}{\partial z}$$ & $$\frac{\partial y'}{\partial v}$$ \\
 $$\frac{\partial z'}{\partial x}$$ & $$\frac{\partial z'}{\partial y}$$ & $$\frac{\partial z'}{\partial z}$$ & $$\frac{\partial z'}{\partial v}$$ \\
 $$\frac{\partial v'}{\partial x}$$ & $$\frac{\partial v'}{\partial y}$$ & $$\frac{\partial v'}{\partial z}$$ & $$\frac{\partial v'}{\partial v}$$ \\

\end{array}
\right)$
\end{center}

The stability around the fixed points depends on the nature of the eigenvalues of this matrix. The point will be stable if all four eigenvalues are negative, unstable if all four eigenvalues are positive, and will be a saddle point if there are both positive and negative eigenvalues.
\section{BARYONIC MATTER AS "CATALYST"}
First of all, we consider the baryonic matter as the ``catalyst'' of dark energy and dark matter, its influence appear in  the interaction term $Q=\alpha H(\rho_{d}+\rho_{m}+\rho_{b})$,  where $\alpha$ is coupling constant. That is to say, dark energy can be converted into dark matter, or in the opposite direction. However, baryonic matter, as the ``catalyst'', it is conserved. In addition to gravitational interaction, the radiation do not interact with the other three components, therefore, it is also  conserved.

In classical Einstein cosmology, the autonomous system (12)-(15) can be written as
\begin{eqnarray}
x^{'}&=&-3 (1-x) x w_d+v x-\alpha  (x+y+z),\\
y^{'}&=&3 x y w_d+v y+\alpha  (x+y+z),\\
z^{'}&=&3 x z w_d+v z,\\
v^{'}&=&3 v x w_d+v^2-v.
\end{eqnarray}
Furthermore, we can obtain the critical points $(x_{c},y_{c},z_{c},v_{c})$ of the autonomous system as follows:
\begin{itemize}
  \item \textbf{1}:~~~
   A=($0$,~~$0$,~~$0$,~~$1$),\\
  \item \textbf{2}:~~~
   B=($\frac{1}{6} \left(3-\frac{\sqrt{3} \sqrt{w_d \left(4 \alpha +3 w_d\right)}}{w_d}\right)$,~~$\frac{1}{6} \left(\frac{\sqrt{3} \sqrt{w_d \left(4 \alpha +3 w_d\right)}}{w_d}+3\right)$,~~$0$,~~$0$),\\
  \item \textbf{3}:~~~
   C=($\frac{1}{6} \left(\frac{\sqrt{3} \sqrt{w_d \left(4 \alpha +3 w_d\right)}}{w_d}+3\right)$,~~$\frac{1}{6} \left(3-\frac{\sqrt{3} \sqrt{w_d \left(4 \alpha +3 w_d\right)}}{w_d}\right)$,~~$0$,~~$0$).

\end{itemize}
In this case, the Jacobian matrix at the fixed point is as follow:
\\

M=$\left(
\begin{array}{cccc}
 $$-\alpha -3 (1-x) w_d+3 x w_d+v$$ &
  $$-\alpha$$ & $$-\alpha$$ & $$x$$ \\
 $$\alpha +3 y w_d$$ & $$\alpha +3 x w_d+v$$ & $$\alpha$$ & $$y$$ \\
 $$3zw_d$$ & $0$ & $$v+3 x w_d$$ & $z$ \\
 $$3 v w_d$$ & $0$ & $0$ & $$3 x w_d+2 v-1$$ \\

\end{array}
\right)$
\\

\begin{table}
\begin{center}
\begin{tabular}{|c||c|c|c|}
  \toprule
  Point & $(x_{c},y_{c},z_{c},v_{c})$ & Eigenvalues & $w_{c}$
\\\hline\hline
  $~$ & ~ & $\lambda_{1}=\lambda_{2}=1$, & ~
  \\
  $A$ & $\left(0,0,0,1\right)$ & $\lambda_{3}=-a-\frac{3 w_d}{2}+1$, & $\frac{1}{3}$
  \\
  $~$ & ~ & $\lambda_{4}=a-\frac{3 w_d}{2}+1$ & ~
  \\\hline
  $~$ & ~ & $\lambda_{1}=\frac{3 w_d}{2}-a$, & ~
  \\
  $B$ & $\left(\frac{1}{2}-\frac{a}{3 w_d},\frac{1}{2}+\frac{a}{3 w_d},0,0\right)$ & $\lambda_{2}=-1+\frac{3 w_d}{2}-a$, & $\frac{w_d}{2}-\frac{a}{3}$
  \\
  $~$ & ~ & $\lambda_{3}=\frac{3 w_d^2-6 a+b}{4 w_d}$  & ~
  \\
  $~$ & ~ & $\lambda_{4}=\frac{3 w_d^2-6 a+b}{4 w_d}$ & ~

  \\\hline
  $~$ & ~ & $\lambda_{1}=a+\frac{3 w_d}{2}$, & ~
  \\
  $C$ & $\left(\frac{1}{2}+\frac{a}{3 w_d},\frac{1}{2}-\frac{a}{3 w_d},0,0\right)$  & $\lambda_{2}=a+\frac{3 w_d}{2}-1$, & $\frac{w_d}{2}+\frac{a}{3}$
  \\
  $~$ & ~ & $\lambda_{3}=\frac{3 w_d^2+6 a-b}{4 w_d}$  & ~
  \\
  $~$ & ~ & $\lambda_{4}=\frac{3 w_d^2+6 a+b}{4 w_d}$ & ~
  \\

  \botrule
\end{tabular}
\end{center}
\caption{where $a=\frac{1}{2} \sqrt{3} \sqrt{w_d \left(4 \alpha +3 w_d\right)}$, $b=\sqrt{6} \sqrt{w_d^3 \left(2 \alpha +\sqrt{3} \sqrt{w_d \left(4 \alpha +3 w_d\right)}+3 w_d\right)}$ The properties of the critical points for the interacting
$Q=\alpha H(\rho_{d}+\rho_{m}+\rho_{b})$.}
\end{table}
One can put the values of the three fixed points into the above matrix respectively and calculate the corresponding eigenvalues, we have the four eigenvalues for each point in Table II. Then we examine the sign of the eigenvalues of the three points. In this literature, our main research is studying the stable point. However, for point A and C, we find they can not be stable point for any value of $\alpha$ and $w_d$, so we don't attend to study their qualities further.  For point B, we find it is stable if one of the following conditions could be satisfied£º
\\
\\$1.\alpha \leq 0$ , $w_d<0 $
\\$2.\alpha >0 $ , $w_d<-\frac{1}{3} (4 \alpha )$ or  $w_d\geq 0$
\\
\\Furthermore, for this stable point B, we have  $\frac{2}{3}<x<\frac{3}{4}$ and $\frac{1}{4}<y<\frac{1}{3}$ and $w<-\frac{1}{3}$ for anyone of the following conditions
\\
\\1. $\frac{1}{4}<\alpha \leq \frac{1}{3}$ , $-\frac{1}{9} (16 \alpha )<w_d<\frac{1}{3 \alpha -3}$
\\2. $\frac{1}{3}<\alpha$ , $-\frac{1}{9} (16 \alpha )<w_d<-\frac{1}{2} (3 \alpha )$
\\
\\ Therefore, in this case of $Q=\alpha H(\rho_{d}+\rho_{m}+\rho_{b})$ and baryonic matter as the "catalyst", there exists one stable attractor B, for this stable attractor point, if the parameters of $\alpha$ and $w_d$ satisfy anyone of above two conditions, we will have $\frac{2}{3}<x<\frac{3}{4}$ and $\frac{1}{4}<y<\frac{1}{3}$ and $w<-\frac{1}{3}$. Then this stable attractor is able to alleviate the cosmological coincidence problem and describe the current stage of accelerated expansion of the universe.
\section{RADIATION AS  "CATALYST"}
In this case, we consider the radiation as the ``catalyst'' of dark energy and dark matter, its influence reflects in  the interaction term $Q=\beta H(\rho_{d}+\rho_{m}+\rho_{r})$,  where $\beta$ is coupling constant. That is to say, dark energy can be converted into dark matter, or in the opposite direction. However, for radiation, as the ``catalyst'', it is conserved. In addition to gravitational interaction, the baryonic matter do not interact with the other three components, therefore it is also conserved.

 In classical Einstein cosmology, the autonomous system (12)-(15) can be written as
\begin{eqnarray}
x^{'}&=&-3 (1-x) x w_d+v x-\beta  (x+y+v),\\
y^{'}&=&3 x y w_d+v y+\beta  (x+y+v),\\
z^{'}&=&3 x z w_d+v z,\\
v^{'}&=&3 v x w_d+v^2-v.
\end{eqnarray}
Furthermore, we can obtain the critical points $(x_{c},y_{c},z_{c},v_{c})$ of the autonomous system as follows:
\begin{itemize}
  \item \textbf{1}:~~~
  A=($0$,~~$0$,~~$1$,~~$0$),\\
  \item \textbf{2}:~~~
  B=($-\frac{\beta }{3 w_d-1}$,~~$-\beta$,~~$0$,~~$\frac{3 \beta  w_d+3 w_d-1}{3 w_d-1}$),\\
  \item \textbf{3}:~~~
  C=($\frac{1}{6} \left(3-\frac{\sqrt{3} \sqrt{w_d \left(4 \beta +3 w_d\right)}}{w_d}\right)$,~~$\frac{1}{6} \left(\frac{\sqrt{3} \sqrt{w_d \left(4 \beta +3 w_d\right)}}{w_d}+3\right)$,~~$0$,~~$0$),\\
\item \textbf{4}:~~~
  D=($\frac{1}{6} \left(\frac{\sqrt{3} \sqrt{w_d \left(4 \beta +3 w_d\right)}}{w_d}+3\right)$,~~$\frac{1}{6} \left(3-\frac{\sqrt{3} \sqrt{w_d \left(4 \beta +3 w_d\right)}}{w_d}\right)$,~~$0$,~~$0$).

\end{itemize}
In this case, the Jacobian matrix at the fixed point is as follow:
\\

M=$\left(
\begin{array}{cccc}
 $$-\beta -3 (1-x) w_d+3 x w_d+v$$ &
  $$-\beta$$ & $0$ & $$x-\beta$$ \\
 $$\beta +3 y w_d$$ & $$\beta +3 x w_d+v$$ & $0$ & $$y+\beta$$ \\
 $$3zw_d$$ & $0$ & $$v+3 x w_d$$ & $z$ \\
 $$3 v w_d$$ & $0$ & $0$ & $$3 x w_d+2 v-1$$ \\

\end{array}
\right)$
\\

\begin{table}
\begin{center}
\begin{tabular}{|c||c|c|c|}
   \toprule
  Point & $(x_{c},y_{c},z_{c},v_{c})$ & Eigenvalues & $w_{c}$
\\\hline\hline
 $~$ & ~ & $\lambda_{1}=-1$, & ~
  \\
  $A$ & $\left(0,0,1,0\right)$  & $\lambda_{2}=0$, & $0$
  \\
  $~$ & ~ & $\lambda_{3}=-c-\frac{3 w_d}{2}$  & ~
  \\
  $~$ & ~ & $\lambda_{4}=c-\frac{3 w_d}{2}$ & ~
 \\\hline
  $~$ & ~ & $\lambda_{1}=\lambda_{2}=1$, & ~
  \\
  $B$ & $\left(\frac{\beta }{1-3 w_d},-\beta,0,\beta +\frac{\beta }{3 w_d-1}+1\right)$ & $\lambda_{3}=\frac{\sqrt{\left(3 w_d-1\right){}^2}}{1-3 w_d}c-\frac{3 w_d}{2}+1$, & $\frac{1}{3}$
  \\
  $~$ & ~ & $\lambda_{4}=\frac{\sqrt{\left(1-3 w_d\right){}^2}}{3 w_d-1}c-\frac{3 w_d}{2}+1$ & ~
  \\\hline
  $~$ & ~ & $\lambda_{1}=\frac{3 w_d}{2}-c$, & ~
  \\
  $C$ & $\left(\frac{1}{2}-\frac{c}{3 w_d},\frac{1}{2}+\frac{c}{3 w_d},0,0\right)$ & $\lambda_{2}=-1+\frac{3 w_d}{2}-c$, & $\frac{w_d}{2}-\frac{c}{3}$
  \\
  $~$ & ~ & $\lambda_{3}=-\frac{-3 w_d^2+6 c+d}{4 w_d}$  & ~
  \\
  $~$ & ~ & $\lambda_{4}=\frac{3 w_d^2-6 c+d}{4 w_d}$ & ~

  \\\hline
  $~$ & ~ & $\lambda_{1}=c+\frac{3 w_d}{2}$, & ~
  \\
  $D$ & $\left(\frac{1}{2}+\frac{c}{3 w_d},\frac{1}{2}-\frac{c}{3 w_d},0,0\right)$  & $\lambda_{2}=c+\frac{3 w_d}{2}-1$, & $\frac{w_d}{2}+\frac{c}{3}$
  \\
  $~$ & ~ & $\lambda_{3}=\frac{3 w_d^2+6 c-d}{4 w_d}$  & ~
  \\
  $~$ & ~ & $\lambda_{4}=\frac{3 w_d^2+6 c+d}{4 w_d}$ & ~
  \\

  \botrule
\end{tabular}
\end{center}
\caption{where  $c=\frac{1}{2} \sqrt{3} \sqrt{w_d \left(4 \beta +3 w_d\right)}$, $d=\sqrt{6} \sqrt{w_d^3 \left(2 \beta +\sqrt{3} \sqrt{w_d \left(4 \beta +3 w_d\right)}+3 w_d\right)}$, The properties of the critical points for the interacting $Q=\beta H(\rho_{d}+\rho_{m}+\rho_{r})$.}
\end{table}
One can put the values of the three fixed points into the above matrix respectively and calculate the corresponding eigenvalues, we have the four eigenvalues for each point in Table II. Then we examine the sign of the eigenvalues of the four points. For point A, B and D, we find they can not be stable point for any value of $\beta$ and $w_d$, so it is not necessary to study their qualities further. For point C, we find it is stable if one of the following conditions could be satisfied£º
\\
\\$1.\beta \leq 0$ , $w_d<0 $
\\$2.\beta >0 $ , $w_d<-\frac{1}{3} (4 \beta )$ or  $w_d\geq 0$
\\
\\Furthermore, for point C, we have  $\frac{2}{3}<x<\frac{3}{4}$ and $\frac{1}{4}<y<\frac{1}{3}$ and $w<-\frac{1}{3}$ for anyone of the following conditions
\\
\\1. $\frac{1}{4}<\beta \leq \frac{1}{3}$ , $-\frac{1}{9} (16 \beta )<w_d<\frac{1}{3 \beta -3}$
\\2. $\frac{1}{3}<\beta$, $-\frac{1}{9} (16 \beta )<w_d<-\frac{1}{2} (3 \beta )$
\\
\\So, in this case of $Q=\beta H(\rho_{d}+\rho_{m}+\rho_{b})$ and considering radiation as the ``catalyst'', there exists one stable attractor C, and for this stable attractor point C, if the parameters of $\beta$ and $w_d$ satisfy anyone of above two conditions, we will have $\frac{2}{3}<x<\frac{3}{4}$ and $\frac{1}{4}<y<\frac{1}{3}$ and $w<-\frac{1}{3}$. Then this stable attractor would have $ \frac{\Omega _{\text{DE}}}{\Omega _{\text{DM}}}$ of the order one and can describe the current stage of accelerated expansion of the Universe.

Until now, we find a very interesting thing which is that the result of this case is very similar to the above case's, even the expression of the stable attractor point and the value range of the two parameters are common to each other. That means no matter considering baryonic matter or the radiation as the ``catalyst'', if they affect the transfer of dark sector like the way we proposed and the two parameters correspondingly take the same value from the available value range, then they can finally get the same stable attractor point.

\section{BOHT RADIATION AND THE BARYONIC MATTER AS  "CATALYST"}
Finally, we consider both the baryonic matter and radiation as the "catalyst" of dark energy and dark matter. Their influence are embodied in the interaction term $Q=\gamma H(\rho_{d}+\rho_{m}+\rho_{b}+\rho_{r})$,  where $\gamma$ is coupling constant. That is to say, just like the former two cases, dark energy can be converted into dark matter, or in the opposite direction. However, for baryonic matter and radiation, as the ``catalyst'', they are always conserved respectively.

 In classical Einstein cosmology, the autonomous system (12)-(15) can be written as
\begin{eqnarray}
x^{'}&=&-\gamma -3 (1-x) x w_{d}+v x,\\
y^{'}&=&\gamma +3 x y w_{d}+v y,\\
z^{'}&=&3 x z w_{d}+v z,\\
v^{'}&=&3 x z w_{d}+v^{2} -v.
\end{eqnarray}
Furthermore, we can obtain the critical points $(x_{c},y_{c},z_{c},v_{c})$ of the autonomous system as follows:
\begin{itemize}
  \item \textbf{1}:~~~
  A=($\frac{1}{6} \left(\frac{\sqrt{3} \sqrt{w_{d} \left(4 \gamma +3 w_{d}\right)}}{w_{d}}+3\right)$,~~$\frac{1}{6} \left(3-\frac{\sqrt{3} \sqrt{w_{d} \left(4 \gamma +3 w_{d}\right)}}{w_{d}}\right)$,~~$0$,~~$0$),\\
  \item \textbf{2}:~~~
  B=($\frac{1}{6} \left(3-\frac{\sqrt{3} \sqrt{w_{d} \left(4 \gamma +3 w_{d}\right)}}{w_{d}}\right)$,~~$\frac{1}{6} \left(\frac{\sqrt{3} \sqrt{w_{d} \left(4 \gamma +3 w_{d}\right)}}{w_{d}}+3\right)$,~~$0$,~~$0$),\\
  \item \textbf{3}:~~~
  C=($\frac{\gamma }{1-3 w_{d}}$,~~$-\gamma$,~~$0$,~~$\gamma +\frac{\gamma }{3 w_{d}-1}+1$).
\end{itemize}
 in this case, the Jacobian matrix at the fixed point is as follow:
\\

M=$\left(
\begin{array}{cccc}
 $$v-3 (1-x) w_d+3 x w_d $$& $0$ & $0$ & $x$ \\
 $$3 y w_d$$ & $$v+3 x w_d$$ & $0$ &$ y $\\
 $$3 z w_d$$ & $0$ & $$v+3 x w_d $$& $z$ \\
 $$3 v w_d$$ & $0$ & $0$ & $$2 v+3 x w_d-1$$ \\
\end{array}
\right)$
\\

We can put the values of the three fixed points into the above matrix respectively and calculate the corresponding eigenvalues. the results are in Table III. In this coupled case, We list the four eigenvalues for each point in Table III. We examine the sign of the eigenvalues of the three points. Then we find that point A and C can not be stable point for any value of $\gamma$ and $w_d$, however, we find point B is stable if one of the following conditions could be satisfied£º
 \\
\\1. $\gamma \leq 0$ , $w_d<0$
\\2. $\gamma >0$ , $ w_d<-\frac{1}{3} (4 \gamma )$ or $w_d>0$
\\
\\Furthermore, for point B, we have  $\frac{2}{3}<x<\frac{3}{4}$ and $\frac{1}{4}<y<\frac{1}{3}$ and $w<-\frac{1}{3}$ for anyone of the following conditions
\\
\\1. $\frac{1}{4}<\gamma \leq \frac{1}{3}$ , $-\frac{1}{9} (16 \gamma )<w_d<\frac{1}{3 \gamma -3}$
\\2. $\gamma >\frac{1}{3}$ , $ -\frac{1}{9} (16 \gamma )<w_d<-\frac{1}{2} (3 \gamma )$
\\
\\
So this point can describe the current accelerated expansion of the Universe and have the $ \frac{\Omega _{\text{DE}}}{\Omega _{\text{DM}}}$ of the order for some values of $\gamma$ and $w_{d}$.

Besides, we can see that the expression of the stable attractor point B and the value range of the two parameters are the same to the first two cases', in addition, if we consider dark energy interaction with dark matter in the absence of ``catalyst'', let us assume $Q=\eta H(\rho_{d}+\rho_{m})$, then we find the expression of the stable attractor point and the value range of the two parameters are also the same to the first three cases'. That means that if we request $\alpha=\beta=\gamma=\eta$ in the common desirable range and take the same $w_{d}$, the Universe will finally reach the same stable point. This is reasonable, like catalyst in the chemical reaction, although they can influence the reaction rate of the materials, they do not change the equilibrium state. baryonic matter or radiation or both of them as the ``catalyst'' could affect the process to obtain the stable attractor point because for every situation, its autonomous system is different from the others. In this paper,we don't attend to discuss the the details influenced by catalysts. however, regardless of the presence of the ``catalyst'', in the end, it is the inevitable result that they will reach the same stable attractor point which is not related to the  ``catalyst'', and the stable attractor point can lead a accelerated Universe and has the  $ \frac{\Omega _{\text{DE}}}{\Omega _{\text{DM}}}$ of the order one.

  We have to admit that, from our results, we are not sure which way is the true evolution of the Universe, and we know $\Omega _{\text{b}}$ and $\Omega _{\text{r}}$  are not strictly zero in today's Universe. However, the distributing condition of cosmic material may be consistent to our results with the evolution of time. That is to say, in the future, we will have $ \frac{\Omega _{\text{DE}}}{\Omega _{\text{DM}}}$ of order one and both $\Omega _{\text{b}}$ and $\Omega _{\text{r}}$ approach to zero.
\begin{table}
\begin{center}
\begin{tabular}{|c||c|c|c|}
  \toprule
  Point & $(x_{c},y_{c},z_{c},v_{c})$ & Eigenvalues & $w_{c}$
\\\hline\hline
  $~$ & ~ & $\lambda_{1}=\lambda_{2}=\frac{3 w_d}{2}-e$, & ~
  \\
  $A$ & $\left(\frac{1}{2}-\frac{e}{3 w_d},\frac{1}{2}+\frac{e}{3 w_d},0,0\right)$ & $\lambda_{3}=-1+\frac{3 w_d}{2}-e$, & $\frac{w_d}{2}-\frac{e}{3}$
  \\
  $~$ & ~ & $\lambda_{4}=-2 e$ & ~
  \\\hline
  $~$ & ~ & $\lambda_{1}=\lambda_{2}=\frac{3 w_d}{2}+e$, & ~
  \\
  $B$ & $\left(\frac{1}{2}+\frac{e}{3 w_d},\frac{1}{2}-\frac{e}{3 w_d},0,0\right)$ & $\lambda_{3}=-1+\frac{3 w_d}{2}+e$, & $\frac{w_d}{2}+\frac{e}{3}$
  \\
  $~$ & ~ & $\lambda_{4}=2 e$ & ~
  \\\hline
  $~$ & ~ & $\lambda_{1}=\lambda_{2}=1$, & ~
  \\
  $C$ & $\left(\frac{\gamma }{1-3 w_{d}},-\gamma ,0,\gamma +\frac{\gamma }{3 w_{d}-1}+1\right)$ &$\lambda_{3}=\frac{\sqrt{\left(1-3 w_d\right){}^2}}{1-3 w_d}e-\frac{3 w_d}{2}+1$,& $ \frac{1}{3}$
  \\
  $~$ & ~ & $\lambda_{4}=\frac{\sqrt{\left(1-3 w_d\right){}^2}}{3 w_d-1}e-\frac{3 w_d}{2}+1$ & ~

  \\
  \botrule
\end{tabular}
\end{center}
\caption{where $e=\frac{1}{2} \sqrt{3} \sqrt{w_d \left(4 \gamma +3 w_d\right)}$ the properties of the critical points for the interacting
$Q=\gamma H(\rho_{d}+\rho_{m}+\rho_{b}+\rho_{r})$.}
\end{table}
\section{Conclusions}
In this paper, we have mainly examined three catalyst type of interactions between dark energy and dark matter. For every case, we  have got a stable attractor point that can lead our present accelerated Universe and had $ \frac{\Omega _{\text{DE}}}{\Omega _{\text{DM}}}$ of order one. Besides, we also have discussed the value range of the parameters and found that for all the cases including the case of no ``catalyst'', the limitation of the two parameters and the expression of the stable attractor point remain unchanged. That implies if we give all the coupling constants the same value and take a certain state parameter $w_d$ of dark energy, our Universe will finally reach the same stable point. It is a necessary consequence that we treat baryonic matter or radiation or both of them as the ``catalyst''. Therefore, the significance of our paper is that we proposed a new relationship between the normal matter(baryonic and radiation) and dark sector that can alleviate the coincidence problem.

\end{document}